\documentclass[doublecol]{epl2}
\usepackage{graphicx}
\usepackage{amsmath,bm}

\title{Zero-temperature spin-glass freezing in self-organized arrays of Co nanoparticles}
\author{R. L\'opez-Ruiz\inst{1} \and F. Luis$^{\ast}$ \inst{1}\and J. Ses\'e\inst{2} \and J. Bartolom\'e\inst{1} \and C.
Deranlot\inst{3} \and F. Petroff\inst{3}} \shortauthor{R. L. L\'opez
\etal}

\institute{
  \inst{1} Instituto de Ciencia de Materiales de Arag\'on
  - CSIC-Universidad de Zaragoza, 50009 Zaragoza, Spain,
  and Departamento de F\'{\i}sica de la Materia Condensada, Universidad de Zaragoza, 50009 Zaragoza, Spain\\
  \inst{2} Instituto de Nanociencia de Arag\'on, Universidad de Zaragoza,
  and Departamento de F\'{\i}sica de la Materia Condensada, Universidad de
  Zaragoza, 50009 Zaragoza, Spain\\
  \inst{3} Unit\'e Mixte de Physique CNRS/Thales - Route D\'epartementale 128, 91767 Palaiseau Cedex,
  France, and Universit\'e Paris-Sud - 91405 Orsay Cedex, France\\}


\abstract{We study, by means of magnetic susceptibility and magnetic
aging experiments, the nature of the glassy magnetic dynamics in
arrays of Co nanoparticles, self-organized in $N$ layers from $N=1$
(two-dimensional limit) up to $N=20$ (three-dimensional limit). We
find no qualitative differences between the magnetic responses
measured in these two limits, in spite of the fact that no
spin-glass phase is expected above $T=0$ in two dimensions. More
specifically, all the phenomena (critical slowing down, flattening
of the field-cooled magnetization below the blocking temperature and
the magnetic memory induced by aging) that are usually associated
with this phase look qualitatively the same for two-dimensional and
three-dimensional arrays. The activated scaling law that is typical
of systems undergoing a phase transition at zero temperature
accounts well for the critical slowing down of the dc and ac
susceptibilities of all samples. Our data show also that dynamical
magnetic correlations achieved by aging a nanoparticle array below
its superparamagnetic blocking temperature extend mainly to nearest
neighbors. Our experiments suggest that the glassy magnetic dynamics
of these nanoparticle arrays is associated with a zero-temperature
spin-glass transition.}

\begin{document}

\maketitle


\section{Introduction}
Dense arrays of magnetic nanoparticles contain the physical
ingredients of spin-glasses \cite{Mydosh93}. Disorder in the
positions and orientations of the particles leads to disorder and
frustration of the dipolar interactions, usually dominant, between
their magnetic moments. In contrast with "canonical" spin-glasses,
the slow magnetic relaxation introduced by interactions coexists
with the slow magnetization reversal associated with the high
anisotropy energy barriers. Many experiments performed on dense
nanoparticulate materials show phenomena, such as magnetic aging
\cite{Jonsson95} and the slowing down of the ac susceptibility
\cite{Djurberg97}, which are {\em typical} of spin-glasses
\cite{Mulder81,Souletie85,Lundgren83,Chamberlin84,Granberg88,Lefloch92,Jonason98,Jonsson02}.
However, some of these phenomena are not {\em exclusive} of the
spin-glass phase \cite{Mattsson93,Schins93}. The question is, then,
whether real nanoparticulate materials show a true (super)spin-glass
phase.

Experimental studies are often hindered by the lack of control over
the sample parameters that determine the nature and strength of
dipolar interactions, such as interparticle distances, spatial
organization, etc. This usually makes it difficult to know {\em a
priori} if a particular system is expected to show a spin-glass
phase. Perhaps the most clear-cut situation to discuss the existence
of a phase transition and its experimental manifestations is offered
by the study of a single layer of nanoparticles. In contrast with
three-dimensional systems \cite{Palassini99}, it is generally
accepted \cite{Young83,Dekker88} that the transition temperature
$T_{\rm g}$ vanishes in two-dimensions. Results of tempered Monte
Carlo simulations seem to confirm the same conclusion also for Ising
spins coupled by dipolar interactions \cite{Fernandez08}.

Based on these considerations, our work was aimed to elucidate the
nature of the glassy magnetic dynamics, i.e. whether it is
associated with a superspin-glass phase at $T_{\rm g}>0$ or if, by
contrast, $T_{\rm g}=0$, in self-organized nanoparticle arrays. For
this, we compare results obtained on very well-characterized three-
and two-dimensional arrays of Co nanospheres. Previous experiments
reveal that the superparamagnetic blocking temperature $T_{\rm b}$,
defined as the temperature of the in phase $\chi^{\prime}$ ac
susceptibility cusp, increases as additional layers are deposited on
a two-dimensional sample \cite{Luis02a,Luis02b}. Since the number of
layers modifies the number of nearest neighbors in the nanoparticle
array, that result indicates that dipolar interactions slow down the
magnetic relaxation processes. In the present study, we have
investigated how the number of layers modifies the critical slowing
down and the magnetic aging, properties that are usually associated
with the spin glass behavior \cite{Petracic06}. Our results show
that no qualitative changes in these quantities occur as the
two-dimensional limit is approached. The control over the number of
layers and their separation has also enabled us to directly probe
the magnetic correlation length and show that it is mainly
restricted to a first shell of nearest neighbors and, in any case,
shorter than what would be expected for a conventional spin glass.


\section{Experimental details}
Samples made of $N$ layers of Co nanoparticles with average diameter
$D \simeq 2.6$ nm were prepared by the sequential sputtering of $N =
1, 2, 3, 4, 5, 7, 10, 15,$ and $20$ Co and Al$_{2}$O$_{3}$ layers on
silicon substrates \cite{Maurice99,Babonneau00,Luis02a,Luis02b}. The
particle's shape and average size (thus also the average magnetic
moment $\mu_{\rm p}$ per particle), as well as the width of the size
distribution ($\sigma_{D} = 0.26 D$) are approximately independent
of $N$ \cite{Luis02b}. Nanoparticles deposited on adjacent layers
tend to self-organize in a structure that resembles a closed-packed
hexagonal lattice of nanospheres \cite{Babonneau00}. The separation
between the Co layers is determined by the thickness $t_{{\rm
Al}_{2} {\rm O}_{3}} = 3$ nm of the alumina layer. Nearest neighbors
separations are $d_{\rm nn,\|} \simeq 4.6$ nm, within a given layer,
and $d_{\rm nn,\bot} \simeq 4.2$ nm, between adjacent layers. They
correspond to dipolar energies $E_{\rm dip} = \mu_{\rm p}^{2}/d_{\rm
nn}^3 \approx 13$ K and $17$ K, respectively. As described in
\cite{Luis03}, the anisotropy energy barrier $U_{0}$ for the
magnetization reversal was estimated from ac susceptibility
experiments performed under sufficiently strong magnetic fields,
which dominate over dipolar interactions. This method gives $U_{0}
\simeq 430$ K. In the same way, we estimate an attempt time
$\tau_{0} \sim 10^{-13}$ s, of the same order of that found for
samples of very small Co nanoparticles ($D \sim 1$ nm), prepared by
the same technique \cite{Luis02a}, for which interactions are
expected to become neglibible. A multilayer with $N=20$ layers but a
larger interlayer separation $t_{{\rm Al}_{2} {\rm O}_{3}} = 10$ nm,
and thus also a much smaller interlayer $E_{\rm dip} \approx 1.6$ K,
was prepared under identical experimental conditions.

Ac susceptibility and magnetization measurements were performed with
a commercial SQUID magnetometer. Samples were rectangular plates
with approximate dimensions $9 \times 3 \times 0.5$ mm$^{3}$. Ac and
dc magnetic fields were parallel to the plane of the sample to
minimize demagnetizing effects. In our study of aging
\cite{Lundgren83,Chamberlin84}, we measured the time-dependent
relaxation of the zero-field cooled (ZFC) magnetization on samples
aged, at zero field, for a time $t_{\rm w}$ at temperatures $T_{\rm
w} < T_{\rm b}$. In addition, we employed a different method which
consists on measuring magnetization curves (zero-field and field
cooled (FC), and remanence) using the waiting time protocol
described in \cite{Mathieu01a,Sasaki05}.


\section{Results and discussion}
A typical method to characterize the spin-glass behavior is by
measuring the frequency-dependent ac magnetic susceptibility
\cite{Mulder81,Souletie85,Djurberg97}. At any fixed frequency
$\omega$ we define a characteristic relaxation time $\tau_{\rm c}$
such that $\omega \tau_{\rm c}(T)=1$ at $T = T_{\rm b}(\omega)$. For
spin glasses $\tau_{\rm c}$ diverges at $T_{\rm g}$ according to a
power law, reflecting the growth of magnetic correlations
\cite{Mulder81}

\begin{equation}
\tau_{\rm c} = \tau_{\ast}|1-T/T_{\rm{g}}|^{-z \nu} \label{tauSG}
\end{equation}

\begin{figure}
\onefigure[width=8cm]{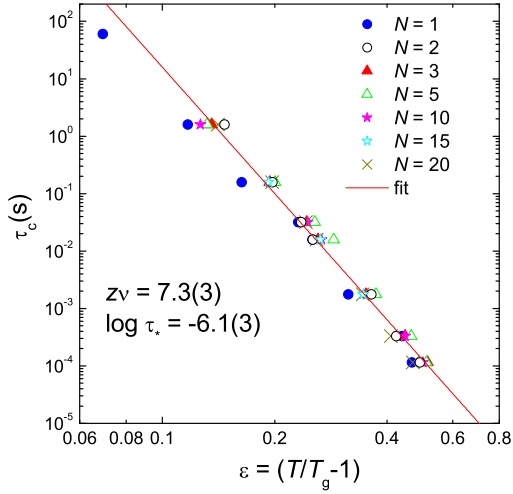} \caption{(Color online).
Critical slowing down of the characteristic relaxation time
extracted from ac susceptibility experiments. Results are shown for
samples with varying number of layers $N$.} \label{slowingdown}
\end{figure}

In Fig. \ref{slowingdown}, we plot $\tau_{\rm c}$  versus the
reduced temperature for $N$ ranging from $1$ to $20$ layers. The
experimental data are compatible with a critical slowing down of the
magnetization dynamics at a finite $T_{\rm g}$. In order to limit
the number of fitting parameters, we took $T_{\rm g}$, for each
sample, as the temperature of the ZFC susceptibility cusp (i.e.
equal to the $T_{\rm b}$ corresponding to a typical timescale of the
order of $170$ s). The microscopic time scale $\tau_{\ast}$ and the
dynamical critical exponent $z \nu$ are found to be nearly the same
for all samples. The critical exponent is close to typical values
found for spin glasses \cite{Souletie85} as well as for some
nanoparticulate materials \cite{Wandersman08}. The present results
are remarkable because it is generally believed that $T_{\rm g} = 0$
in two dimensions. Notice however that, as often happens with plots
of this type obtained for nanoparticles
\cite{Djurberg97,Wandersman08,Petracic02}, experiments do not
explore the close vicinity of the critical region. For this reason,
the data are relatively easy to fit. Fits of similar quality can be
obtained by scaling all $T_{\rm g}$'s by a factor in between $1$ and
$0.75$. The characteristic $\tau_{\ast}$ increases then from about
$10^{-6}$ s to $10^{-4}$ s while, at the same time, the exponent
$z\nu$ increases from $7.3$ to $14$. In fact, if one wishes to
include also in the analysis the freezing temperature extracted from
ZFC susceptibility data (getting closer to $T_{\rm g}$), the best
fits with Eq. (\ref{tauSG}) are obtained then for the largest $z\nu$
and $\tau_{\ast}$ values (thus also for the lowest $T_{\rm g}$).
Such large $z\nu$ values are not uncommon in systems of magnetic
nanoparticles \cite{Djurberg97,Petracic02} but they are
significantly larger than what it is expected for a canonical spin
glass phase transition (of the order of $z\nu=7$ \cite{Ogielsky85}).

\begin{figure}
\onefigure[width=7cm]{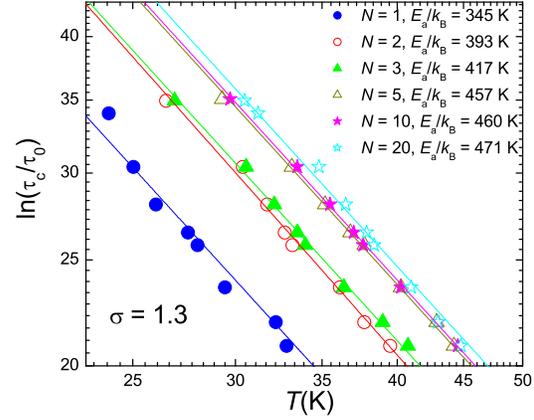} \caption{(Color online).
Log-log plot showing the variation with temperature of
$\ln(\tau_{\rm c}/\tau_{0})$, where $\tau_{\rm c}$ is a
characteristic relaxation time extracted from ac susceptibility data
and $\tau_{0} = 10^{-13}$ s. The lines are fits of the law
$\tau_{\rm c} = \tau_{0} \exp{\left( E_{\rm a}/k_{\rm B} T
\right)^{\sigma}}$, characteristic of a spin-glass transition at
$T_{\rm g}=0$} \label{Tg0}
\end{figure}

An alternative theoretical framework to describe the
frequency-dependent susceptibility, which seems very appropriate in
the case of a layered material with a markedly two-dimensional
character, is the activated dynamics characteristic of glassy
systems undergoing a phase transition at $T_{\rm g}=0$
\cite{Mulder81,Dekker88}. In the latter situation, the critical
slowing down of $\tau_{\rm c}$ obeys the following expression

\begin{equation}
\tau_{\rm c} = \tau_{0} \exp{\left( E_{\rm a}/k_{\rm B} T
\right)^{\sigma}} \label{activatedtauc}
\end{equation}

\noindent where $E_{\rm a}$ is an effective activation energy and
$\sigma$ is a critical exponent. As it is shown in Fig. \ref{Tg0},
we find a good agreement with our data, including also the
temperature of the ZFC magnetization cusp, for $\sigma=1.3$ (to be
compared with $\sigma = 3.2$ found for $2-D$ spin-glasses
\cite{Dekker88}) and $E_{\rm a}$ gradually increasing with the
number of layers from $345$ K up to $471$ K. From these
frequency-dependent susceptibility experiments, we conclude that the
nature of the slow magnetic dynamics of two-dimensional (i.e. with
$N$ equal to or close to unity) and three-dimensional (with large
$N$) nanoparticle arrays is the same. The description based on a
zero-temperature phase transition is appealing, because it is
consistent with the behavior expected for a single layer. By
themselves, however, these experiments cannot discriminate between
the two alternatives, i.e., whether the underlying physics
corresponds to the existence of a second-order phase transition at a
finite $T_{\rm g}$ or if, by contrast, $T_{\rm g}=0$.

\begin{figure}
\onefigure[width=8cm]{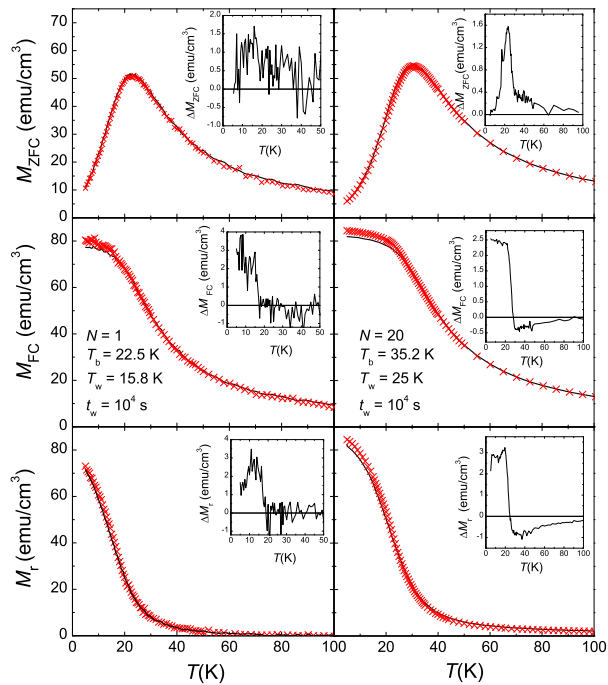} \caption{ (Color online). ZFC, FC
and thermoremanence curves measured with (red crosses) and without
(black solid line) pause. The insets show absolute values of $\Delta
M_{\rm ZFC}$, $\Delta M_{\rm FC}$ and $\Delta M_{\rm r}$. Left:
sample with $N=1$; $T_{\rm w}$ was $15.8$ K. Right: sample with
$N=20$; $T_{\rm w}$ was $25$ K. The waiting time was $10^{4}$ s and
the applied magnetic field was $10$ Oe.} \label{aging}
\end{figure}

Aging experiments can shed some light and help deciding between
these two alternatives, since they probe how dynamical magnetic
correlations grow with time \cite{Lundgren83,Granberg88,Jonsson02}.
We have carried out two different experiments, which measure the
magnetic memory effects associated with the aging of the sample at a
given temperature. In the first of these, the quantity of interest
is the difference $\Delta M = M - M_{\rm{w}}$ between the
magnetizations (ZFC, FC or remanent) measured after cooling the
sample without or with a pause at an intermediate temperature
$T_{\rm w}< T_{\rm b}$ \cite{Mathieu01a}. Results measured for
$t_{\rm w} = 10^{4}$ s are shown in Fig. \ref{aging}. $\Delta M_{\rm
ZFC}$ shows a peak centered near $T_{\rm w}$. If $T_{\rm w}$ is
varied, the peak shifts accordingly. In addition, the relationship
\cite{Mathieu01a} $\Delta M_{\rm FC} = \Delta M_{\rm r}+\Delta
M_{\rm ZFC}$ is fulfilled, showing that they are associated with the
aging of the sample at $T_{\rm w}$ and not with experimental
artifacts.

        %
        %

Figure \ref{aging} compares results obtained on a single layer $N=1$
with those measured on a multilayer made of $N=20$ layers. The aging
was performed at $T_{\rm w} = 0.7 T_{\rm b}$ for the two samples.
Besides the obvious difference in the signal-to-noise ratios, they
look qualitatively the same. The maximum in $\Delta M_{\rm ZFC}$ vs
$T$ is just about $25$ \% larger in the case of the multilayer. A
first conclusion is, therefore, that the magnetic memory induced by
aging a nanoparticle array does not show any abrupt change as the
two-dimensional ($2D$) limit is approached. Notice also that, as we
have seen with the critical slowing down, the analogy is not
restricted to aging. The FC curves measured on the two samples show
also the same degree of ``flattening'' below $T_{\rm b}$, a property
that has been considered as a signature of the superspin glass phase
\cite{Sasaki05}.

\begin{figure}
\onefigure[width=8cm]{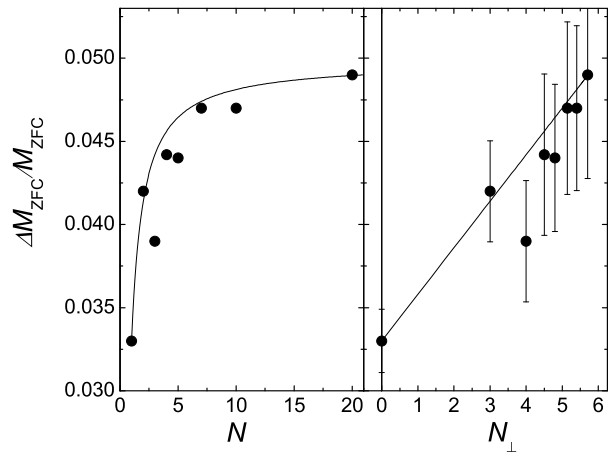} \caption{(Color online). Left:
Variation with the number $N$ of layers of the memory $\Delta M_{\rm
ZFC}/M_{\rm ZFC}$ measured after aging the sample for $t_{\rm w} =
10^{4}$ s at $T_{\rm w} = 15.8$ K ($\bullet$). Right: Same data as a
function of the average number of nearest neighbors
$N_{\bot}=6(N-1)/N$ that a nanoparticle has in adjacent layers
\cite{Luis02b}. The lines represent $0.033+0.015 N_{\bot}$.}
\label{MaxHolevsN}
\end{figure}

\begin{figure}
\onefigure[width=8cm]{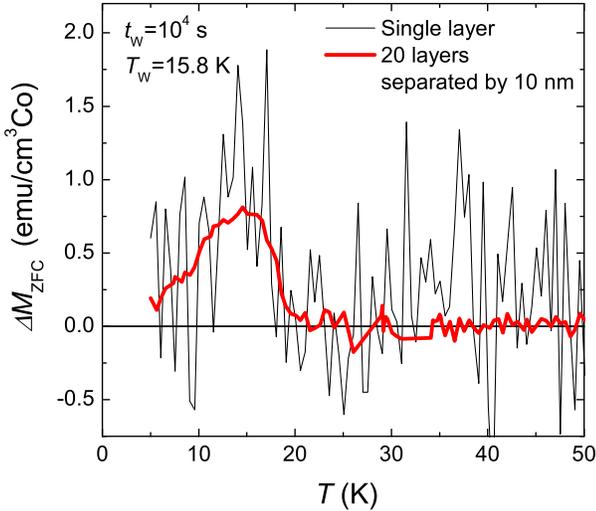} \caption{(Color online). Magnetic
memory for a single layer of 2.6 nm particles (black thin line)
compared with data obtained for a $N = 20$ multilayer in which the
interlayer separation is $t_{\rm Al_{2} \rm O_{3}} = 10$ nm (red
thick line).} \label{Aging4}
\end{figure}

By gradually changing the number of layers $N$ we can
study how magnetic correlations grow. In Fig. \ref{MaxHolevsN}, we
show, as a function of $N$, the relative amplitude of the magnetic
memory effect $\Delta M_{\rm ZFC} /M_{\rm ZFC}$ measured after aging
at $T_{\rm w}=15.8$ K for $t_{\rm w} = 10^{4}$ s. Within the droplet
picture of the spin-glass phase \cite{Fisher88}, this quantity is
connected with the size that domains of correlated spins attain
after time $t_{\rm w}$ \cite{Jonsson02}. We see that $\Delta M_{\rm
ZFC}/M_{\rm ZFC}$ increases rapidly when one or two layers are added
to a two-dimensional sample, nearly saturating as $N$ increases
further. The right-hand panel of Fig. \ref{MaxHolevsN} shows that,
within the relatively large experimental uncertainties, $\Delta
M_{\rm ZFC}/M_{\rm ZFC}$ is approximately proportional to the
increase in the average number of nearest neighbors $N_{\perp} =
6(N-1)/N$ that is associated with the addition of extra layers. The
same linear dependence was also found for the blocking temperature
\cite{Luis02b}. This behavior suggests that the enhancement in the
amplitude of the magnetic memory is provided mainly by correlations
with the first one or two nearest layers. Another result suggesting
that magnetic correlations remain rather short-ranged is shown in
Fig. \ref{Aging4}. There, we compare the magnetic memory $\Delta
M_{\rm ZFC}$ obtained for a single layer of $2.6$ nm particles with
that obtained for a $N = 20$ multilayer in which the interlayer
separation is $t_{\rm Al_{2} \rm O_{3}} = 10$ nm, i.e. more than
twice $d_{{\rm nn},\|} \sim 4.6$ nm. Within their respective
experimental uncertainties, these two quantities are found to be the
same. Also $T_{\rm b}$ and other quantities agree. It seems then
that no measurable magnetic correlations are established between
layers of nanoparticles located $10$ nm far form each other.

\begin{figure}
\onefigure[width=8cm]{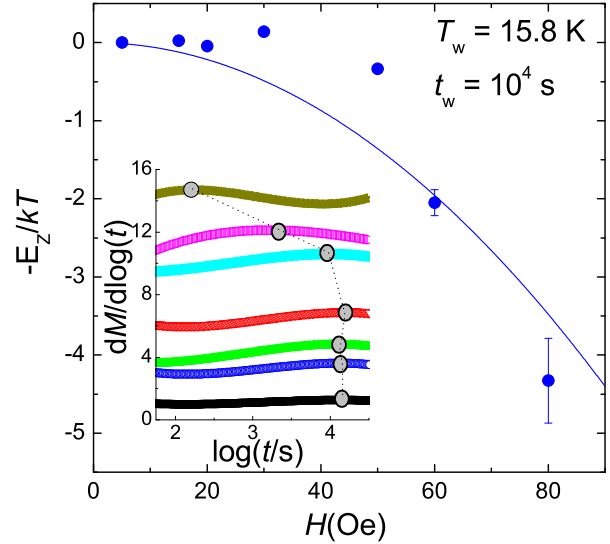} \caption{(Color online). Zeeman
energies estimated from the relaxation of ZFC magnetization curves
measured after aging the sample at zero field and at $T_{\rm w} =
15.8$ K for a waiting time $t_{\rm W} = 10^{4}$ s, before the
application of a magnetic field $H$. The inset shows the time
dependence of the magnetic viscosity, defined as $\partial M/
\partial \log(t)$ with $M$ being the sample's magnetization,
measured for (from bottom to top curves) $H=5$ $15$, $20$, $30$,
$50$, $60$, and $80$ Oe. This quantity shows maxima, marked by grey
dots, at the effective age of the system that decreases with
increasing $H$. The dotted line is a guide to the eye.}
\label{EzvsH}
\end{figure}

We also studied the effects of aging using a different experimental
method, which enables a more quantitative determination of
magnetic correlation lengths. For this, we measured the magnetic
relaxation of the ZFC magnetization of a $N=15$ multilayer at
$T_{\rm w}=15.8$ K. The sample was first cooled from $100$ K to
$T_{\rm w}$ in zero field. After aging the sample for $t_{\rm w} =
10^{4}$ s, a magnetic field $H$ was applied and the ensuing
magnetization measured as a function of time. This method has been
applied to estimate the number of correlated spins in spin glasses
\cite{Joh99,Bert04}, and recently applied also to investigate the
slow dynamics of frozen ferrofluids \cite{Wandersman08}. Its basic
idea is as follows. During the waiting time $t_{\rm w}$, at zero
field, magnetic correlations between nanoparticles grow
\cite{Lundgren83}. Typical free energy barriers $\Delta (t_{\rm w})$
for the flip of $N_{\rm s}(t_{\rm w})$ correlated spins increase
also with the age $t_{\rm w}$ of the system. This growth of
dynamical correlations reflects itself in the appearance of a
maximum in the relaxation rate, defined as $\partial M_{\rm
ZFC}/\partial \log(t)$, when the experimental time $t$ approaches
the age of the system $t_{\rm w}^{\rm eff} \sim t_{\rm w}$ (see Fig.
\ref{EzvsH}). A magnetic field $H$ reduces the free energy barriers,
from its zero-field value to $\Delta (t_{\rm w})-E_{\rm Z}\left[
H,N_{\rm s}(t_{\rm w}) \right]$, where $E_{\rm Z} \left[ H,N_{\rm
s}(t_{\rm w}) \right] = \mu \left[ H,N_{\rm s}(t_{\rm w}) \right] H$
and $\mu$ is the magnetic moment of a "drop" of $N_{\rm s}$
correlated spins. The energy shift induced by this Zeeman term
effectively reduces the "age" of the system according to
\begin{equation}
t_{\rm w}^{\rm eff}(H)=t_{\rm w}^{\rm eff}(H=0) \exp
\left\{-\frac{E_{\rm Z}\left[ H,N_{\rm s}(t_{\rm w}) \right]}{k_{\rm
B}T} \right\} \label{ageeff}
\end{equation}
\noindent therefore shifting the relaxation rate maximum towards
shorter times with increasing $H$, as it is indeed observed
experimentally (Fig. \ref{EzvsH}). From a series of experiments
performed at different fields, ranging from $5$ Oe up to $100$ Oe,
we have extracted the Zeeman energy $E_{\rm Z}\left[ H,N_{\rm
s}(t_{\rm w}) \right]$, which we plot in the main panel of Fig.
\ref{EzvsH}. The field dependence of this energy can be fitted using
a quadratic function of $H$, compatible with the following
expression

\begin{equation}
E_{\rm Z}\left[ H,N_{\rm s}(t_{\rm w}) \right] = N_{\rm s}(t_{\rm
w})\chi_{\rm ZFC}H^{2} \label{EZ}
\end{equation}

\noindent which was found to agree also with experiments performed
on frozen ferrofluids \cite{Wandersman08}. Inserting in Eq.
(\ref{EZ}) the measured FC susceptibility per particle $\chi_{\rm
FC}$, we estimate the number of correlated spins $N_{\rm s}$ to be
approximately $17$, i.e., rather close to the average number of
nearest neighbors ($12$) in a multilayer of nanoparticles
\cite{Luis02b}.

Our experimental findings suggest therefore that magnetic
correlations achieved after aging the sample at low $T$ extend
mainly to nearest neighbors. The detailed characterization of our
samples enables us to make a quantitative comparison of the present
results with predictions for the growth of correlations in
spin-glasses. Theoretical considerations as well as experiments
support the idea that correlations grow approximately as a power law
of time \cite{Wandersman08,Joh99,Komori99}:

\begin{equation}
\xi(t^{*},T_{\rm w})/d_{\rm nn} \sim \left( t^{*} \right)^{\alpha
(T_{\rm w})} \label{correlation}
\end{equation}

\noindent where $d_{\rm nn} \sim 4.4$ nm is the distance to nearest
neighbors, $t^{*} = t_{\rm w}/ \tau(T_{\rm w})$ is a dimensionless
timescale, the exponent $\alpha = 0.17 (T_{\rm w}/T_{\rm g})$, and
$\tau$ is the relaxation time of individual spins at the given
temperature. We have estimated $\tau = \tau_{0}\exp{\left( U/k_{\rm
B}T \right)}$ using parameters estimated, as described above, for
the noninteracting case: $\tau_{0} \sim 10^{-13}$ s and $U \simeq
430$ K \cite{Luis03}. For $T_{\rm w} = 15.8$ K and $t_{\rm
w}=10^{4}$ s, Eq. \ref{correlation} gives $4.2 d_{\rm nn} < \xi < 7
d_{\rm nn}$, i.e., between $19$ and $31$ nm for a single layer and
$3d_{\rm nn} < \xi < 4.2 d_{\rm nn}$ ($13-19$ nm) for a multilayer.
The upper and lower limits of $\xi$ correspond to, respectively, the
lower and upper limits of the freezing temperatures $T_{\rm g}$ that
are compatible with the ac susceptibility experiments described
above. Our magnetic memory experiments point to significantly
shorter correlation lengths $\xi \sim 4.4$ nm.


\section{Conclusions}
The central result of the present study is that we observe the same
magnetic memory and critical slowing down in two-dimensional
nanoparticle arrays, as well as in multilayers, suggesting that the
underlying physical behavior is also the same. The slowing down of
the ac (and dc) susceptibility curves measured on all these samples
can, in fact, be accounted for using the activated law
[Eq.(\ref{activatedtauc})] that is typical of two-dimensional
spin-glasses. These results suggest, therefore, that the glassy
magnetic dynamics observed in these materials is associated with a
phase transition occurring at $T_{\rm g}=0$, rather than with a
conventional spin-glass transition with a finite $T_{\rm g}$. This
conclusion is supported by the results of magnetic memory
experiments, which show that dynamical magnetic correlations are
rather short ranged and, in any case, shorter than expected for
canonical spin glasses.

As mentioned in the introductory section above, the
observed facts disagree with the prediction, derived from Monte
Carlo simulations, that Ising-like spins interacting via
dipole-dipole interactions should undergo a spin-glass transition
below a finite temperature\cite{Fernandez08}. Establishing the
origin for this discrepancy is beyond the scope of the present work.
Here, we content ourselves with discussing possible deviations of
real materials from the ideal conditions set by such models. In our
opinion, an important aspect to be considered is the, unavoidable,
distribution in particle sizes. In our multilayers, the distribution
is properly described by a Gaussian function of width $\sigma_{D}
\simeq 0.7$ nm, which is equivalent to roughly $\pm 1$ atomic layer
and provides an indication of the good homogeneity of these samples.
This narrow size distribution leads, however, to an extremely large
dispersion in the relaxation times $\tau$ associated with the
magnetic anisotropy of the nanoparticles. Using the parameters $U$
and $\tau_{0}$ given above (see also the second reference in
\cite{Luis02b} for further details), it follows that intrinsic
timescales separated by more than $13$ orders of magnitude can
coexist at temperatures near or below $T_{\rm b}$. A possible
consequence of this enormous dispersion is the following. Smaller,
and therefore faster, relaxing nanoparticles are able to immediately
react to spin flips of larger (slower) ones, minimizing their mutual
interaction energy. We have previously shown that this effect
accounts for the modification of the average relaxation times by
interactions, at temperatures close to $T_{\rm b}$ \cite{Luis02b}.
We might speculate with the possibility that the disorder in
relaxation times also hinders the growth of magnetic correlations at
lower temperatures. For instance, the formation of negatively
polarized magnetic clouds surrounding the largest nanoparticles can
screen dipolar interactions between them. Clearly, further
theoretical studies that include effects of disorder and
nonequilibrium dynamics are required to clarify the nature of the
collective magnetic response in nanoparticle arrays.


\begin{acknowledgments}
This work was partly funded under grants MAT08/1077,
MAT2009-13977-C03 and "Molecular Nanoscience" (CSD2007-00010) from
Spanish MICINN and PI091/08 "NABISUP" from DGA.
\end{acknowledgments}
$^{\ast}$ Corresponding author. Email: fluis@unizar.es


\end{document}